\documentclass{PoS}

\usepackage{amsmath}
\usepackage{amsfonts}
\usepackage{amssymb}

\usepackage{mathtools}
\usepackage{graphicx}
\usepackage{subfigure}

\newcommand{\eq}[1]{\begin{equation}
                     \begin{split} #1 \end{split}
                     \end{equation}}
\newcommand{\ov}{\overline}                   
\newcommand{\tri}{\hspace{-3.5pt}\vartriangle\hspace{-3.5pt}}
\newcommand{\bom}[1]{\fboxsep2mm\fbox{
           $ \displaystyle{ #1} $}}
           
\title{Non-geometric fluxes and non-associative geometry}

\ShortTitle{Non-geometric fluxes and non-associative geometry}

\author{Erik Plauschinn \\
         Institute for Theoretical Physics and Spinoza Institute \\
Utrecht University \\ 3508 TD  Utrecht, The Netherlands \\
        E-mail: \email{e.plauschinn@uu.nl}}

\abstract{
In these proceedings, we discuss non-commutativity in closed string theory. In analogy to the open-string sector, for closed strings we first motivate a cyclic double commutator to be evaluated for backgrounds with geometric or non-geometric fluxes. A non-trivial result for such an expression  indicates a non-associative structure. Second, we define a conformal field theory at linear order in background fluxes and compute correlation functions therein. From these we  motivate a tri-product which  captures non-commutative and non-associative effects.
}

\FullConference{Proceedings of the Corfu Summer Institute 2011 School and Workshops on Elementary Particle Physics and Gravity\\
September 4-18, 2011\\
Corfu, Greece \\[5mm]
{\rm \footnotesize Report numbers: SPIN-12/09 and ITP-UU-12/10}\\[-5mm]}

\begin{document}


\section{Introduction and Motivation}

One of the remarkable features of string theory is that it provides a framework to treat gauge theories and gravity in a unified way, and which is expected to be complete in the ultra-violet (UV). 
Concerning gauge theories in field theory, it is known how to deal with (certain classes of) thereof in the UV via the procedure of renormalization. For gravity on the other hand, that question is understood to a far lesser degree. Although, one may expect that a theory of quantum gravity is related to a space-time which is non-commutative.
In view of this expectation, also  string theory should feature a non-commutative behavior.

However, non-commutativity in string theory was first discovered for D-branes which correspond to the gauge theory sector. More concretely, for an open string ending on a D-brane endowed with a gauge flux, the commutator of two open-string coordinates on the brane is generically non-vanishing \cite{Connes:1997cr,Chu:1998qz,Schomerus:1999ug}. 
Moreover, correlation functions of vertex operators on the D-brane 
also indicate a non-commutative structure which is encoded in a phase factor. 
Constructing then an effective action out of these correlators, it turns out that the phase factor can be incorporated via a Moyal-Weyl star product between the fields. One therefore obtains a non-commutative gauge theory on the D-brane \cite{Seiberg:1999vs}.

For the gravity sector in string theory, non-commutativity appears to be harder to obtain (for earlier work on that subject see for instance \cite{Dasgupta:2003us}).
But, given the results for the open string, a guideline to study  the closed string  may be to parallel the  discussion of the former. 
Let us therefore highlight the following three points:
\begin{enumerate}
 \itemsep0pt

\item For the open string, as will be explained in more detail below, the non-commutativity parameter is related to the gauge flux on the D-brane. Therefore, in the closed-string sector we should consider backgrounds with non-trivial fluxes.

\item A quantity which clearly shows the non-commutative behavior for the open string is the commutator of two coordinates on the D-brane. For the closed string, we thus seek for a similar expression.

\item Correlation functions of open-string vertex operators  exhibited a non-commutative behavior, which is related  to the Moyal-Weyl star product. Therefore, also in the closed-string sector we 
should compute correlation functions and try to extract a non-commutative product.

\end{enumerate}


\paragraph{Background fluxes}

Let us continue with a brief discussion of background fluxes. As just mentioned, for the open string the non-commutativity parameter is related to the gauge flux on the D-brane.  For the closed string, the authors in \cite{Blumenhagen:2010hj} thus considered a background with non-vanishing  $H$-flux realized via Wess-Zumino-Witten model \cite{Witten:1983ar}. Other flux backgrounds with geometric flux $f$ and non-geometric fluxes $Q$ and $R$ can  be obtained by applying successive T-dualities \cite{Dasgupta:1999ss,Shelton:2005cf}
\eq{
  H_{xyz} \;\xleftrightarrow{\;\; T_{z}\;\;}\;
   f_{xy}{}^{z} \;\xleftrightarrow{\;\; T_{y}\;\;}\;
  Q_{x}{}^{yz} \;\xleftrightarrow{\;\; T_{x}\;\;}\;
  R^{xyz} \; .
}
For the setting in \cite{Blumenhagen:2010hj}, the most interesting situations were non-vanishing $H$- and $R$-fluxes, where the latter  is expected to be related to a non-associative structure \cite{Bouwknegt:2004ap,Bouwknegt:2004tr}.
However, the question of non-commutativity in closed string theory was also studied in \cite{Lust:2010iy}, where the cases of $H$-flux and geometric flux were analyzed in the framework of doubled geometry, leading to similar findings as in  \cite{Blumenhagen:2010hj}.

\pagebreak

To summarize, in order to investigate non-commutativity for the closed string, recent work suggests to consider backgrounds with non-vanishing $H$-flux as well as non-vanishing geometric or non-geometric fluxes.


\paragraph{Three-bracket}

We now turn to the question of how to identify a suitable expression displaying non-commutative behavior.
\begin{figure}[t]
\hspace{\stretch{1}}
\subfigure[Open string disc correlator with two vertex operators]{
\includegraphics[width=0.26\textwidth]{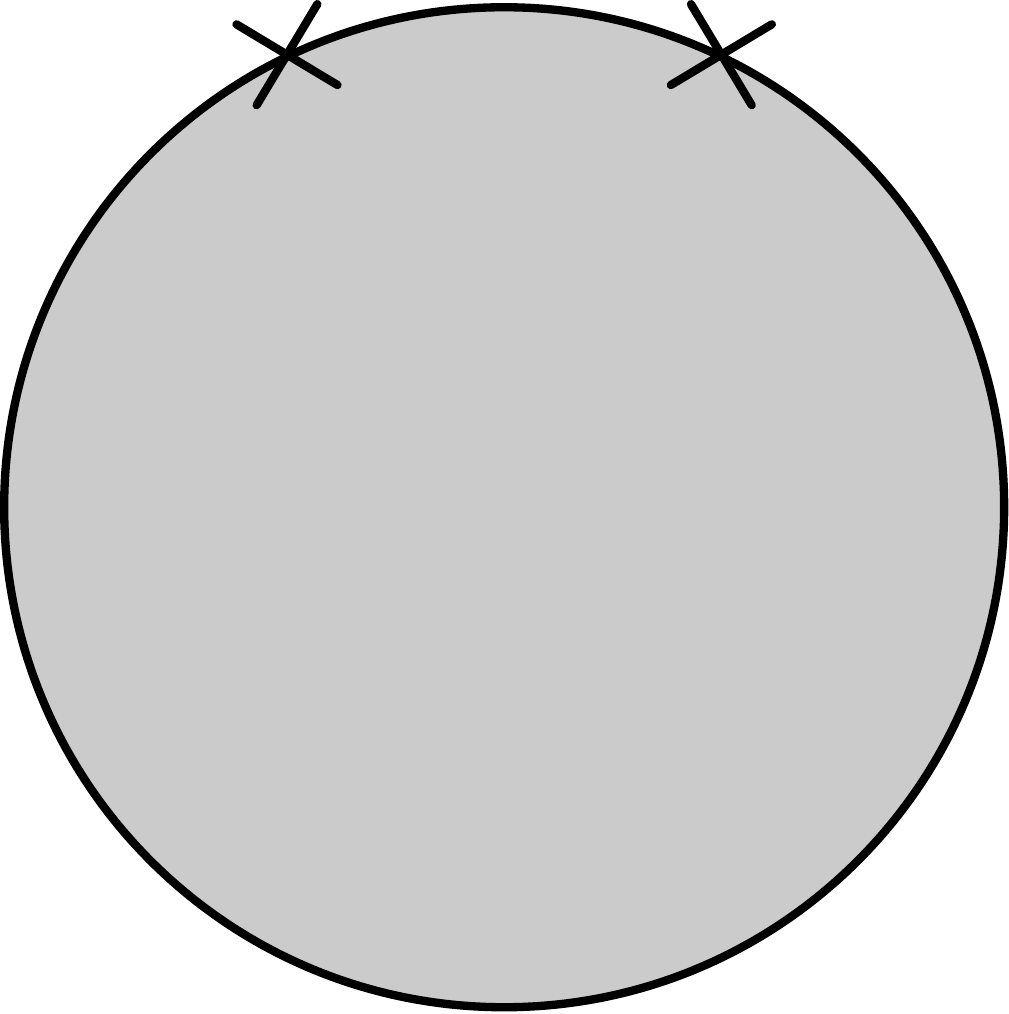}
}
\hspace{\stretch{3}}
\subfigure[Closed string correlator on the sphere with two vertex operators]{
\includegraphics[width=0.26\textwidth]{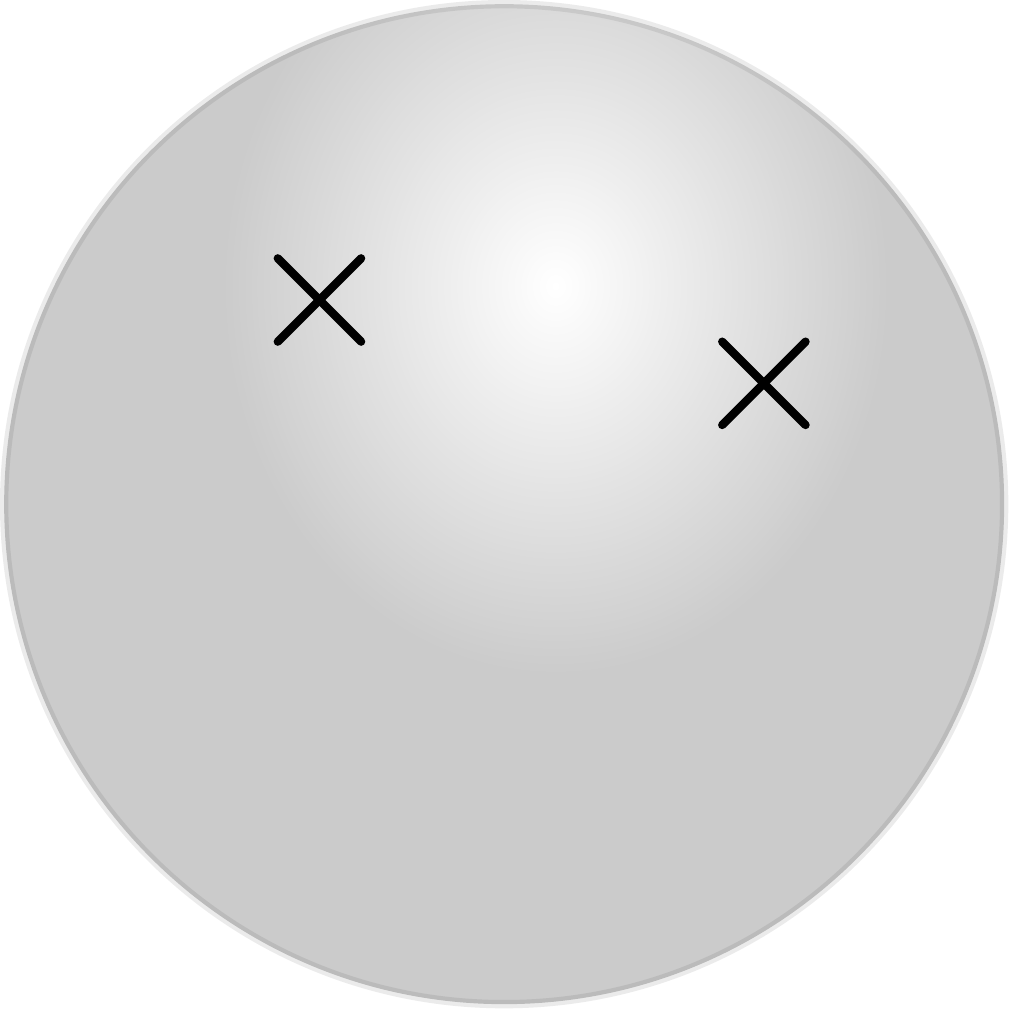}
}
\hspace{\stretch{3}}
\subfigure[Closed string correlator on the sphere with three vertex operators]{
\includegraphics[width=0.26\textwidth]{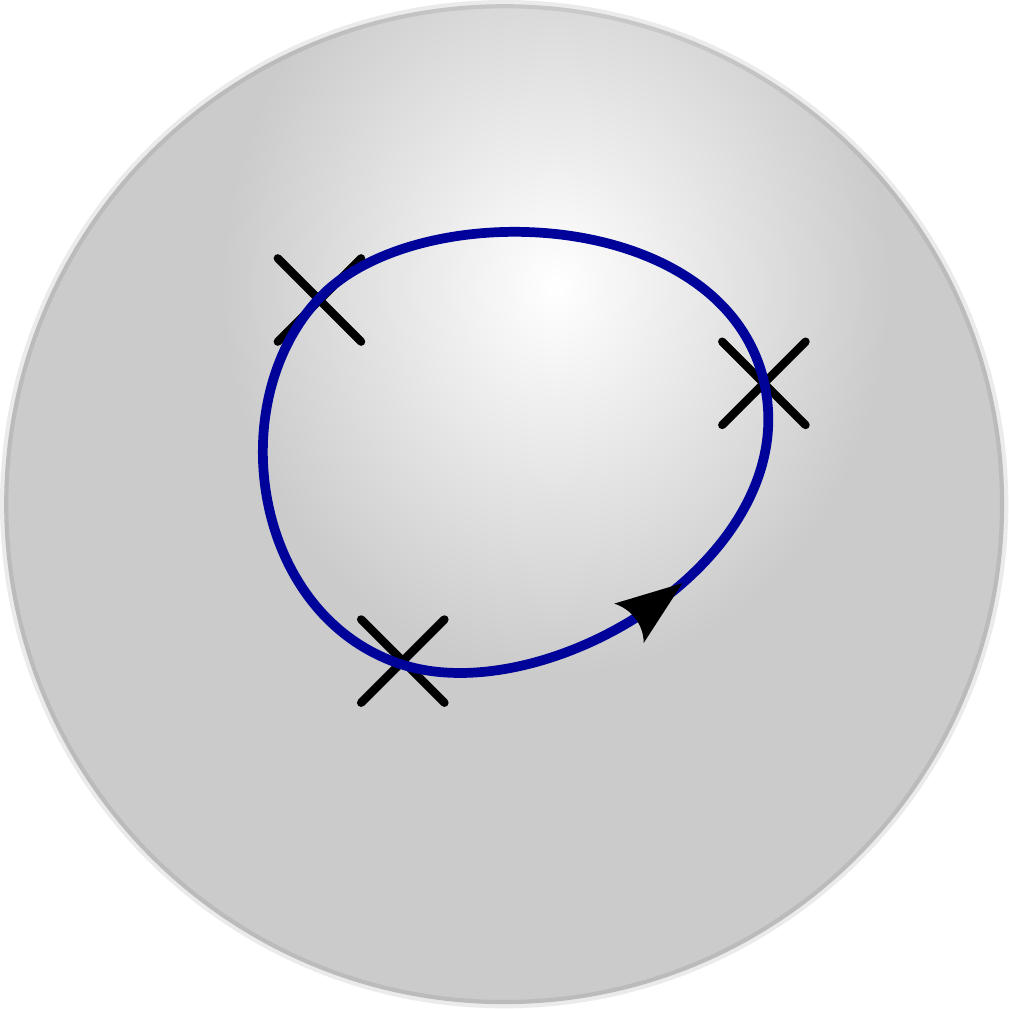}
}
\hspace{\stretch{1}}
\caption{Open string disc diagram (a) with two vertex operators, for which an ordering can be defined (locally). For two vertex operators on the sphere (b), no such ordering can be defined. For three vertex operators on the sphere (c), a line connecting the three insertion points has an orientation (locally).\label{fig_01}}
\end{figure}
For the open string, the origin of the non-commutativity  can be explained heuristically in the following way. To compute the two-point function on the D-brane, as illustrated in figure \ref{fig_01}(a), vertex operators are inserted at the boundary of a disc. Provided there is a quantity sensitive to the ordering of these operators, such as the gauge flux on the brane, the operators do not commute.
For the closed string, the situation is different. Here, the world-sheet is a sphere rather than a disc, and  vertex operators are inserted in the bulk of the sphere. As illustrated in figure \ref{fig_01}(b), no ordering of \emph{two} operators on the sphere can be defined. However, when inserting \emph{three} vertex operators and connecting them through a closed line, as shown in figure \ref{fig_01}(c), a different orientation of the line may be detected by a quantity sensitive to such a change.

This heuristic observation suggests that in order to observe  non-commutativity for the closed string, one should consider three instead of two objects. In terms of commutators of closed string coordinates, a natural guess for such an expression is 
\begin{equation}
\label{jacobiid}
\begin{split}
  \bigl[X^\mu,X^\nu,X^\rho\bigr]:=
  \lim_{\sigma_i\to \sigma}\; \bigl[ [X^\mu (\sigma_1,\tau),
    X^\nu(\sigma_2,\tau)],X^\rho(\sigma_3,\tau)\bigr] +{\rm cyclic}\;. 
\end{split}
\end{equation}
Note that  if this cyclic double commutator is non-vanishing, it indicates not only a non-commu\-tative but also a non-associative structure (see also \cite{Saemann:2012ex} within the proceedings of this conference).

Another argument supporting  \eqref{jacobiid} is that for two closed-string coordinates $X^{\mu}(\sigma,\tau)$ the single commutator generically depends on the coordinates $\sigma$ and $\tau$ of the world-sheet. Therefore, such a quantity can not be expected to characterize a feature of the target space.


\paragraph{Correlation functions}

Non-commutative behavior  may also be encoded in correlation functions of vertex operators, and so we briefly recall the situation for the open string. With $T_i$ denoting a tachyon vertex operator of the open bosonic string, $p_i$ the corresponding momentum, $\theta$ a flux parameter and $\epsilon(\tau)={\rm sign}(\tau)$, a correlation function of $N$ such operators is found to be
\eq{
  \label{correla}
  \bigl\langle \,T_1\, \ldots T_N \bigr\rangle 
  =\exp\biggl(  i  \sum_{1\le n<m\le N} 
    p_{n,a} \,\theta^{ab}\,   p_{m,b}\,  \epsilon(\tau_n-\tau_m)\biggr)\times
    \bigl\langle \,T_1\, \ldots T_N \bigr\rangle_{\theta=0} \;,
}
which contains an extra phase due to the non-commutative nature of the theory.
One can  then define  an $N$-product $\star_N$ in the following way
\eq{
  \label{Nbracketcon}
  & f_1(x)\, \star_N\,  f_2(x)\, \star_N \ldots \star_N\,  f_N(x) := \\
  &\hspace{70pt}\exp\biggl( i  \sum_{1\leq n< m\leq N}
     \theta^{ab}\,
      \partial^{x_n}_{a}\,\partial^{x_m}_{b}  \biggr)\, f_1(x_1)\, f_2(x_2)\ldots
   f_N(x_N)\Bigr|_{x_1=\ldots = x_N=x} \;,
}
which correctly reproduces the phase appearing in \eqref{correla}. 
Note that these $N$-products are  related to 
the subsequent application of the usual star-product $\star=\star_2$. 
Therefore, by evaluating correlation functions of vertex operators
in open string theory, it is possible to 
derive the Moyal-Weyl product and some of its features \cite{Seiberg:1999vs}.
Moreover, in \cite{Cornalba:2001sm} and \cite{Herbst:2001ai} this product has been studied in the context of the open string, and a non-associative behavior has been found.

By analogy, for the closed string we then also have to  compute correlation functions  and identify  the dependence on the flux. If the latter can be encoded in a phase factor, an analysis along similar lines as for the open string can be performed. 
Although quite ambitious, from these correlation functions a product like the Moyal-Weyl star product may be extracted, and a theory of quantum gravity using such a product may be formulated.


\paragraph{Strategy and outline for these proceedings}

The points alluded to  above suggest that to study non-commutativity for the closed string, a cyclic double commutator in a background with non-vanishing fluxes should be computed. A convenient setting for such an analysis are Wess-Zumino-Witten models, which we employ in section \ref{sec_3bracket} to compute  \eqref{jacobiid} for the coordinates of the closed bosonic string.
Two of the main results  will be that a non-commutative behavior for the closed string can indeed be seen, and that this feature appears already at \emph{linear} order in the flux parameter.

However, since non-commutativity can also be detected in correlation functions, in section~\ref{sec_cft} we define a conformal field theory at \emph{linear} order in the flux and introduce vertex operators. Computing then correlation functions of the latter and studying the phase factor of these expressions, we are led to the definition of a tri-product. One might suspect that this product can  be used to construct a non-commutative theory of gravity, which, however, has not been achieved to date.

As a final remark,  let us mention that these proceedings cover a talk given some time ago and so  further developments and new results  have appeared. In particular, as mentioned above, non-commutativity in closed string theory has also been studied in \cite{Lust:2010iy}, and  examples have been constructed in \cite{Condeescu:2012sp}. A comprehensive summary of the ideas in \cite{Blumenhagen:2010hj} and \cite{Blumenhagen:2011ph}, with a  different emphasis compared to here, can be found in the proceedings \cite{Blumenhagen:2011yv}.
Furthermore, as we pointed out above, backgrounds with $H$-, geometric and non-geometric fluxes are important for the question of closed-string non-commutativity. Recently, such settings have been studied for instance in \cite{Andriot:2011uh} from the point of generalized complex geometry, and in \cite{Andriot:2012wx} in the framework of double field theory (see also \cite{Patalong:2012np}). 
Furthermore, in \cite{Blumenhagen:2012ma} a Palatini formulation of (higher-order) Lovelock gravity has been analyzed, and requiring consistency of the Palatini formalism lead to constraints which can be interpreted as Bianchi identities for geometric and non-geometric fluxes. This analysis will be continued in \cite{deser}.


\clearpage
\section{Closed string non-commutativity}
\label{sec_3bracket}

In this section we approach the question about non-commutativity for closed strings. In the open-string sector, non-commutativity appeared for strings ending on D-branes endowed with a background flux \cite{Connes:1997cr,Chu:1998qz,Schomerus:1999ug}. In the following, we will therefore first briefly review the analysis in the open-string case, and then perform a similar computation for closed strings. This section is based on \cite{Blumenhagen:2010hj}, where further details can be found.


\subsection{Open string non-commutativity}
\label{sec_osnc}

In this subsection, we mainly follow the discussion in \cite{Chu:1998qz}.


\paragraph{Open string computation}

We consider an open string with both endpoints on a D$p$-brane
carrying  constant two-form flux ${\cal F}_{ij}=B_{ij}+F_{ij}$, where $i,j=0,\ldots,p$.
This leads to mixed Neumann-Dirichlet boundary conditions 
longitudinal to the brane, so that the mode expansions for the 
corresponding free bosons read
\begin{equation}
\label{openexpand}
X^i(\sigma,\tau)=x^i_0+ \bigl(\alpha^i_0 \,\tau - \alpha^j_0 {\cal F}_j{}^i
    \sigma\bigr) + \sum_{n\ne 0} \frac{e^{-in\tau}}{n}\Bigl(
        i\,\alpha^i_n \cos (n\sigma) - \alpha^j_n {\cal F}_j{}^i \sin (n\sigma)\Bigr)\,.
\end{equation}
Here we normalized $0\le \sigma\le \pi$, and indices of $\mathcal F_{ij}$ are raised by the inverse metric of the form $\eta^{ij}={\rm diag}\,(-1,+1,\ldots,+1)$. As carried out in \cite{Chu:1998qz}, 
the commutation relations for  the modes appearing in \eqref{openexpand}
can be obtained via canonical quantization.
Using these relations,  the equal-time
commutator is evaluated as
\begin{equation}
\label{commut}
  \bigl[X^i(\sigma_1,\tau),X^j(\sigma_2,\tau)\bigr] =-2i\hspace{1pt}\alpha'  \big(M^{-1}{\cal F}\bigr)^{ij}
  \left[ P(\sigma_1,\sigma_2)  + \sum_{n\ne 0} \frac{\sin
      n(\sigma_1+\sigma_2)}{n}\right] ,
\end{equation}
where $M_{ij}=\delta_{ij}-{\cal F}_i{}^k {\cal F}_{kj}$ and matrix products are understood.
The function $P$ is a continuous 
linear expression in the world-sheet coordinates  $\sigma_i$ of the form
\eq{
\label{open_p}
P(\sigma_1,\sigma_2)= \sigma_1+\sigma_2 -\pi \;,
}
which arises purely from the commutation relations involving the zero modes $x_0^i$ and $\alpha_0^i$. The sum in \eqref{commut} originates  from the oscillator modes $\alpha_n^i$ for $n\neq0$, and can be further evaluated using the Fourier transform
\begin{equation}
\label{gammala}
\gamma(\varphi) =\sum_{n=1}^\infty  \:\frac{\sin (n\varphi)}{ n}=
\left\{ 
\begin{array}{@{\hspace{2pt}}c@{\hspace{15pt}}l}
\frac{1}{2} (\pi-\varphi) &  0<\varphi<2\pi\;, \\[1.5mm]
0&  \varphi=0,2\pi \; . 
\end{array}
\right.
\end{equation}
Using then equations \eqref{open_p} and \eqref{gammala}, we see that for  $0< \sigma_1+\sigma_2 < 2\pi$ 
the two terms in \eqref{commut} cancel.
However, 
on the boundaries $\sigma_1=\sigma_2=0$ and $\sigma_1=\sigma_2=\pi$
one obtains
\begin{equation}
   \label{comm_open_res}
   \bigl[X^i(0,\tau),X^j(0,\tau)\bigr]=-\bigl[X^i(\pi,\tau),X^j(\pi,\tau)\bigr]=
    2\pi i\hspace{1pt}\alpha'  \bigl(M^{-1}{\cal F}\bigr)^{ij}\; .
\end{equation}
In summary, the equal-time, equal-position commutator between two target-space
coordinates $X^i(\sigma,\tau)$ does not vanish along a  D-brane carrying  non-trivial
two-form flux $\mathcal F_{ij}$.


\paragraph{Remarks}

Let us make the following remarks:
\begin{itemize}

\item Even without
knowing  the zero mode contribution  $P(\sigma_1,\sigma_2)$ explicitly, 
we could have \linebreak guessed this function by requiring 
the commutator \eqref{commut} to vanish for generic points on the world-sheet.
In turn, the non-zero result in \eqref{comm_open_res} arises from the boundaries of the open string  due to the discontinuity of $\gamma(\varphi)$ at $\varphi=0\!\mod2\pi$.

\item Since the equal-time, equal-position commutator \eqref{comm_open_res} is
independent of the world-sheet coordinates $\sigma$ and $\tau$, one can indeed
conclude that this world-sheet  computation reveals a feature of the
target space (as probed by an open string).

\end{itemize}


\subsection{Closed string non-commutativity}

For closed strings, we expect non-commutativity to arise for
backgrounds with a non-vanishing flux. As a consequence, the equations of motion require the space to be curved. One of the simplest examples for an exactly solvable configuration are Wess-Zumino-Witten (WZW)  \cite{Witten:1983ar}  models describing compactifications with $H$-flux \cite{Gepner:1986wi}.


\paragraph{Wess-Zumino-Witten models}

Let us therefore start our discussion by considering  the WZW model for the group manifold
$SU(2)$. The corresponding action is given by
\begin{equation}
\label{wzwnaction}
\begin{split}
    S=\hphantom{+}&\:\frac{k}{16\pi} \int_{\partial \Sigma} d^2x \, {\rm Tr}\, \Bigl[ (\partial_\alpha g)(\partial^\alpha g^{-1}) \Bigr]
       \\ 
 -&\:\frac{ik}{24\pi} \int_\Sigma d^3y \, \epsilon^{\tilde\alpha\tilde\beta\tilde\gamma} \:
     {\rm Tr}\,\Bigl[ (g^{-1} \partial_{\tilde\alpha} g)(g^{-1} \partial_{\tilde\beta} g)(g^{-1} \partial_{\tilde\gamma} g) \Bigr]\;,
\end{split}
\end{equation}
where $k\in\mathbb Z^+$ denotes the level and 
 $\Sigma$ is a three-dimensional manifold with boundary $\partial\Sigma$. 
The indices take values $\alpha=1,2$ and $\tilde\alpha,\ldots =1,2,3$, which are raised or lowered by the metrics $h_{\alpha\beta}={\rm diag}(+1,+1)$ and $h_{\tilde \alpha\tilde \beta}={\rm diag}(+1,+1,+1)$, respectively.
Parametrizing an element $g\in SU(2)$ in terms of Hopf coordinates $\eta^i$ as
\begin{equation}
  \label{def_g}
     g=\left(\begin{matrix} e^{i\eta^2} \cos\eta^1 &  e^{i\eta^3} \sin\eta^1 \\
          -e^{-i\eta^3} \sin\eta^1 &  e^{-i\eta^2}
          \cos\eta^1 \end{matrix}\right) \;,
\end{equation}
with $0\le \eta^1\le \pi/2$, $0\le \eta^{2,3}\le 2\pi$, one realizes that the first term in
\eqref{wzwnaction} 
is  a non-linear sigma model 
with target space $S^3$ of radius $R=\sqrt k$, and
the second  term 
corresponds to a background flux proportional to $k$.


\paragraph{Conserved currents and Kac-Moody algebras}

Solving the model \eqref{wzwnaction} directly in terms of Hopf coordinates $\eta^i$
is not easily possible, but  it is well known that the
WZW model  actually is exactly solvable. To see this, we 
introduce a complex coordinate $z=\exp(x^1+ix^2)$ and
define the currents 
\begin{equation}
\label{def_j}
J  = 
J^a \, \frac{\sigma^a}{\sqrt 2}= - k\, \bigl( \partial_z g \bigr)\,   g^{-1}\, ,
\hspace{40pt}
\ov J  = 
\ov J{\vphantom J}^a \, \frac{\sigma^a}{\sqrt 2}= + k\, g^{-1} \bigl( \partial_{\ov z} g \bigr) \;.
\end{equation}
Note that here and in the following, $\sigma^a$ with $a=1,2,3$ are the Pauli matrices and  summation over repeated indices is understood.
From the equation of motion of the WZW model \eqref{wzwnaction} it follows that the currents $J^a$ are holomorphic and that the $\ov J{\vphantom J}^a$ are anti-holomorphic.
Therefore, one  can perform the Laurent expansions 
\begin{equation}
\label{expansion}
    J^a(z)=\sum_{n\in \mathbb Z} j^a_n\, z^{-n-1} \;, \hspace{60pt}
    \ov J{\vphantom J}^a(\ov z)=\sum_{n\in \mathbb Z} \ov j{\vphantom j}^a_n\, \ov z^{-n-1} \;.
\end{equation}
The symmetry transformations of the WZW model then translate into the following commutation relations  for the modes $j^a_n$ and $\ov j{\vphantom j}^a_n$ 
\begin{equation}
  \label{km_alg}
  \begin{split}
   \bigl[j^a_m,j^b_n\bigr]\hspace{0.75pt}&=  i f^{ab}{}_c\, j^c_{m+n} + k\, m\, \delta_{m+n}\,
   \delta^{ab}\; , \\[2mm]
    \bigl[\ov j{\vphantom j}^a_m,\ov j{\vphantom j}^b_n\bigr]&=  i f^{ab}{}_c\, \ov j{\vphantom j}^c_{m+n} 
    + k\, m\, \delta_{m+n}\,  \delta^{ab} \;,
\end{split}
\hspace{60pt}\bigl[j^a_m,\ov j{\vphantom j}^b_n\bigr]=0 \;,
\end{equation}
which define two independent Kac-Moody algebras. Note that the structure constants for $SU(2)$ in our convention read $f^{abc}=\sqrt{2} \,\epsilon^{abc}$, and indices are raised or lowered by $\delta^{ab}$ and $\delta_{ab}$, respectively.
For later reference, let
us then employ the parametrization \eqref{def_g} in \eqref{def_j} and express 
 the two  currents \eqref{expansion}  as follows
\begin{equation}
\label{strombein}
  J^a(z)= - i\sqrt{2}\,k\,  {\rm E}^a{}_i\,  \partial_z \eta^i, \hspace{60pt}
  \ov J{\vphantom J}^a(\ov z)= - i\sqrt{2}\,k\, \ov {\rm E}{\vphantom {\rm E}}^a{}_i\,  \partial_{\ov z} \eta^i\;.
\end{equation}
The matrices ${\rm E}^a{}_i$ and
${\rm \ov E}\vphantom{\rm E}^a{}_i$ are known explicitly (see  \cite{Blumenhagen:2010hj}),
and depend on the coordinates $\eta^i$.


\paragraph{Local coordinates}

So far, we have mainly reviewed the well-known geometry for
the exactly solvable $SU(2)_k$ WZW model.
However,  let us now introduce fields   $X^a(z,\ov z)$ according to 
\begin{equation}
\label{free_coords_X_01}
\begin{split}
  J^a(z)&= - i\,\sqrt{k}\,\partial_z X^a(z,\ov z)= - i\sqrt{2} \,k\,  {\rm
    E}^a{}_i(\vec \eta)\,
  \partial_z \eta^i(z,\ov z)  \;,\\
  \ov J\vphantom{J}^a(z)&= -i\,\sqrt{k}\,\partial_{\ov z} X^a(z,\ov z)= 
  -i\sqrt{2} \,k\,  {\rm \ov E}\vphantom{\rm E}^a{}_i(\vec \eta)\,
  \partial_{\ov z} \eta^i(z,\ov z) \;.
\end{split}
\end{equation}
It is clear that
the $X^a$ do not correspond to  bona fide global coordinates
on $S^3$ since there does not exist a flat metric on $S^3$.
However, as shown in \cite{Braaten:1985is},  if the $X^a(z,\ov z)$ satisfy their  
(free) equations of motion, the $\eta^i(z,\ov z)$ do  so as well.

Next, since the cyclic double commutator \eqref{jacobiid} we are interested in  is a local quantity,  we can imagine to probe 
the geometry around a  point $\vec \eta_0$ on a  three-sphere $S^3$
by a closed string. 
Writing then
\eq{
  \label{free_coords_X_02}
  X^a(z,\ov z)=X^a(z)+\overline{X}\vphantom{X}^a(\overline z) 
}
and using \eqref{free_coords_X_01},
locally we can identify the
left- and right-moving coordinates  as
\begin{equation}
\label{free_coords_X_03}
\begin{split}
  X^a(z)\simeq  \sqrt{2k}\,   {\rm E}^a{}_i(\vec \eta_0)\, \eta^i(z)\;,
  \hspace{60pt}
  \ov X\vphantom{X}^a(\ov z)\simeq  \sqrt{2k}\,  {\rm \ov E}\vphantom{\rm E}^a{}_i (\vec
  \eta_0)\, \ov \eta^i(\ov z)\; .
\end{split}
\end{equation}
The mode expansions of $X^a(z)$ and $\ov X\vphantom{X}^a(\ov z)$ are found 
by  integrating  the expansions of the currents given in
 \eqref{expansion}. In particular, for the holomorphic part we arrive at 
\begin{equation}
\label{xexpand}
   X^a(z)=\frac{i}{\sqrt k}\: x_0^a -  \frac{i}{ \sqrt{k}} \,j^a_0 \log z
    + \frac{i}{  \sqrt{k}}\, \sum_{n\ne 0}
    \frac{j^a_n}{n}\: z^{-n} \;,
\end{equation}
and a similar expression is obtained for the anti-holomorphic part $\ov
X\vphantom{X}^a(\ov z)$. The modes $j_n^a$ in \eqref{xexpand} satisfy the
corresponding Kac-Moody algebra given in \eqref{km_alg}, however, a priori it
is not clear what the precise form of the commutation relations involving
$x_0^a$ is. In the following, we are going to fix this contribution in analogy
to the open string discussed in section \ref{sec_osnc}.


\paragraph{Cyclic double commutator}

Let us consider the cyclic double commutator for the holomorphic part $X^a(z)$ of the  free fields \eqref{free_coords_X_02}
\eq{
  \label{double_123}
  \bigl[X^a(z_1),X^b(z_2),X^c(z_3) \bigr]=
  \bigl[\hspace{1pt} \bigl[ X^a(z_1),X^b(z_2)\bigr]\,,\,X^c(z_3) \bigr] +{\rm cyclic}\;,
}
evaluated at equal times. For our choice of complex coordinates $z_i=\exp(\tau_i + i\sigma_i)$ this implies $|z_1|=|z_2|=|z_3|$, which will always be understood for the expression \eqref{double_123}.
To simplify the following formulae, let us furthermore introduce $\mathbf x^a$, $\mathbf p^a$ and $\mathbf j^a$ as 
\eq{
  \mathbf x^a = \frac{i}{\sqrt k}\: x_0^a \;, \hspace{34pt}
  \mathbf p^a(z) = -  \frac{i}{ \sqrt{k}} \,j^a_0 \log z \;,\hspace{34pt}
  \mathbf j^a(z) =  \frac{i}{  \sqrt{k}}\, \sum_{n\ne 0}
    \frac{j^a_n}{n}\: z^{-n} \;.
}
For the computation of \eqref{double_123}, we first collect all terms involving $\mathbf x^a$ into a so far undetermined function  $\mathcal P^{abc}$
\eq{
  \label{zm_contr_01}
  \mathcal P^{abc}(z_1,z_2,z_3) = 
  \bigl[\hspace{1pt}\mathbf x^a, \mathbf x^b, \mathbf x^c\hspace{1pt} \bigr]
  + \bigl[\hspace{1pt}\mathbf x^a, \mathbf x^b, \, \cdot\, \hspace{1pt} \bigr]
  + \bigl[\hspace{1pt}\mathbf x^a, \, \cdot\, , \,  \cdot\,  \hspace{1pt} \bigr] 
  + \ldots \;.
}
For all other contributions in \eqref{double_123}, we employ the Kac-Moody algebra \eqref{km_alg} of the modes $j^a_n$ as well as the Jacobi identity for the structure constants $f^{ab}{}_c$. 
Apart from \eqref{zm_contr_01}, the only non-vanishing double commutator then reads
\eq{
  \label{double_125}
  \bigl[\hspace{1pt}\mathbf j^a(z_1), \mathbf j^b(z_2), \mathbf j^c(z_3)\hspace{1pt} \bigr]
  = - \frac{f^{abc}}{\sqrt k} \sum_{\substack{n,m\neq 0 \\ n+m\neq 0}} \frac{1}{n\,m} \left( \frac{z_3}{z_1} \right)^n 
  \left( \frac{z_3}{z_2} \right)^m + {\rm cyclic}\;.
}
Remember that  this expression is understood to be evaluated at equal times. For the right-hand side in  \eqref{double_125}, we split the sums in the following way and compute
\eq{
  \label{double_126}
  \Gamma(\sigma_1,\sigma_2,\sigma_3)= & - \sum_{n,m\neq 0} \frac{1}{n\,m} \left( \frac{z_3}{z_1} 
 \right)^n  \left( \frac{z_3}{z_2} \right)^m
 - \sum_{n\neq 0} \frac{1}{n^2} \left( \frac{z_2}{z_1} 
 \right)^n  
  + {\rm cyclic} \\[1.8mm]
 =&  \left\{
  \begin{array}{@{}c@{\hspace{30pt}}l}
   -\pi^2 & \sigma_1=\sigma_2=\sigma_3 \;, \\[1.8mm]
    0
    & {\rm else} \;.
    \end{array}
    \right.
}
Combining  the above results, we arrive at  the equal-time
double commutator
of the holomorphic fields $X^a(z)$ of the form
\eq{
  \label{holdoubleyeah}
  \bigl[X^a(z_1),X^b(z_2),X^c(z_3) \bigr]= \mathcal P^{abc}(z_1,z_2,z_3) 
  + \frac{f^{abc}}{\sqrt k}\:\Gamma(\sigma_1,\sigma_2,\sigma_3) \;.
}
For the computation in the anti-holomorphic sector, we note that the modes $\ov j\vphantom{j}^a_n$ satisfy the same Kac-Moody algebra as $j^a_n$. Furthermore, we have $\ov z_i = e^{\tau_i - i \sigma_i}$ and so we only need to replace $\sigma_i\to -\sigma_i$ in the result \eqref{holdoubleyeah} for the holomorphic sector. However, observe that the function $\Gamma$ is invariant under that substitution.
Therefore, the result for the full equal-time double commutator reads
\eq{
\label{doubleyeah}
  &\bigl[X^a(z_1,\ov z_1),X^b(z_2,\ov z_2),X^c(z_3,\ov z_3) \bigr]  \\
  &\hspace{80pt}=\mathcal P^{abc}(z_1,z_2,z_3) 
  + \ov{\mathcal{ P}}\vphantom{\mathcal P}^{abc}(\ov z_1,\ov z_2,\ov z_3) 
  + 2\:\frac{f^{abc}}{\sqrt k}\:\Gamma(\sigma_1,\sigma_2,\sigma_3) \;. 
}
It is now tempting to follow the same logic as for the open-string
computation. That is, we fix the unknown contribution $\mathcal P + \ov{\mathcal{P}}$ of the zero modes $x_0^a$ and $\ov x\vphantom{x}^a_0$ by:
\begin{center}
\label{assum_1}
\begin{tabular}{@{}p{70pt}p{320pt}@{}}
{\it Assumption}\hspace{1.5pt}: & The zero mode contribution $\mathcal P + \ov{\mathcal{P}}$ 
is {\it continuous}; and for the three points $z_i$ not all equal, the equal-time double commutator has to vanish.
\end{tabular}
\end{center}
More concretely, this assumption means that 
\begin{equation}
\label{qqsecond}
  \mathcal P^{abc}(z_1,z_2,z_3) 
  + \ov{\mathcal{ P}}\vphantom{\mathcal P}^{abc}(\ov z_1,\ov z_2,\ov z_3) 
  = 0\;.
\end{equation}
Using then \eqref{double_126} and \eqref{qqsecond} in \eqref{doubleyeah}, we arrive at the following result
\begin{equation}
  \label{result}
   \bigl[X^a,X^b,X^c \bigr] 
   := \lim_{z_i\to z} \bigl[X^a(z_1,\ov z_1),X^b(z_2,\ov z_2),X^c(z_3,\ov z_3) \bigr] 
   =  - \:\frac{2\pi^2}{\sqrt k}\:  f^{abc}\; .
\end{equation}
Therefore, pursuing the same reasoning as for the open string,
we are led to the intriguing result that the fields
$X^a$  satisfy a non-vanishing three-bracket, 
where the right-hand side is constant and proportional to
the $SU(2)$ structure constants $f^{abc}$.


\paragraph*{Summary and remarks}

Let us make the following remarks:
\begin{itemize}

\item In order to get a better understanding of the expression \eqref{result}, we mention that for a fundamental product $x^i \bullet x^j$ one 
can define a three-bracket as
\begin{equation}
\begin{split}
   \bigl[x^1,x^2,x^3\bigr] = \sum_{\sigma\in P_3} {\rm sign}(\sigma) \Bigl(
   \bigl(x^{\sigma(1)}\bullet x^{\sigma(2)}\bigr)\bullet x^{\sigma(3)} 
  - x^{\sigma(1)}\bullet \bigl(x^{\sigma(2)}\bullet x^{\sigma(3)}\bigr) \Bigr)\;,
\end{split}
\end{equation}
being the  completely anti-symmetrized 
associator of this $\bullet\,$-product. 
For an associative product this expression vanishes, and so
the non-vanishing result \eqref{result} indicates
both a non-commutative and non-associative (NCA) structure.

\item The equal-time, equal-position  double commutator is independent
of the world-sheet coordinates. Thus, it is  expected to
reflect a property of the target space (as probed by a closed string).

\item Recalling that the radius of the three-sphere is  $R=\sqrt k$, we
  realize that in the large radius limit   $R\to \infty$ the NCA effect vanishes.

\item We  also computed the single commutator 
$\lim_{z_i\to z} [X^a(z_1,\ov z_1),X^b(z_2,\ov z_2) ]$ 
and found it to be dependent on the world-sheet coordinates. We therefore conclude
that the fundamental, well-defined target space structure is a 
three-bracket.

\item As explained in more detail in  \cite{Blumenhagen:2010hj}, in contrast to 
the Hopf coordinates $\eta^i$ the fields $X^a(z,\ov z)$ in \eqref{free_coords_X_02} are 
not proper coordinates on the sphere.
Therefore, the structure constants $f^{abc}$ appearing on the right-hand side in \eqref{result} have to be interpreted properly.
In particular, they should be interpreted as a non-geometric $R$-flux.

\end{itemize}
To summarize, denoting the flux-parameter by $\theta^{abc}$, in the case of non-geometric $R$-flux the equal-time, equal-position cyclic double commutator indicates a non-associative structure and reads
\begin{equation}
  \label{result2}
   \bom{
   \bigl[X^a,X^b,X^c \bigr] 
   := \lim_{z_i\to z} \Bigl[X^a(z_1,\ov z_1),\bigl[X^b(z_2,\ov
       z_2),X^c(z_3,\ov z_3) \bigr]\Bigr] +{\rm cycl.} 
   =   \theta^{abc}\; .
   }
\end{equation}


\section{Correlation functions and tri-product}
\label{sec_cft}

In this section, we first define a conformal field theory (CFT) for the closed-string sector which captures
effects at linear order in the background flux.
Then, we compute correlation functions of vertex operators therein, and extract a non-commutative product. This part is based on  \cite{Blumenhagen:2011ph}, where  more details  can be found.


\subsection{Structure of the conformal field theory}


\paragraph{Setting}

Here, we do not start from an exactly solvable
WZW model and then take a local limit, but rather
from a flat background with flux.
More concretely, our framework is that of a flat space with constant $H$-flux
and dilaton 
which is to be considered as part of a full bosonic string theory construction. 
The metric and the flux are specified by 
\eq{ 
  \label{setup_01}
   ds^2=\sum_{a=1}^N \bigl(dX^a\bigr)^2, \hspace{50pt} 
   H=  \frac{2}{{\alpha'}^2}\,  \theta_{abc}\,  dX^a\wedge dX^b\wedge dX^c\; ,
} 
where in the following we focus mostly  on  $N=3$.
A closed string moving in this background can be described by a  sigma-model. With $\Sigma$ denoting the world-sheet of the closed string, its action reads
\eq{
  \label{action_730178}
  \mathcal S=\frac{1}{2\pi\alpha'}\int_\Sigma d^2 z\, \bigl(\,
  g_{ab} + B_{ab} \bigr)\, \partial X^a \,\ov\partial X^b
  \;,
}
where the metric for our particular situation is given by $g_{ab} = \delta_{ab}$, and for the $B$-field we choose a gauge in which $B_{ab} = \frac{1}{3}  H_{abc}\, X^c$.

Let us point out that already at lowest order in $\alpha'$, the background given by \eqref{setup_01}
is not a solution
to the string equations of motion. 
In particular, the beta-functional for the graviton
\eq{
  \label{targeteom}
  \beta_{ab}^G  =  \alpha' R_{ab}-\frac{\alpha' }{4}\: H_{a}{}^{cd}\, H_{bcd}  
  +2\alpha' \nabla_a\nabla_b\Phi
    +O({\alpha'}^2) 
}
does not vanish for \eqref{setup_01} in the case of a constant dilaton $\Phi$. 
Only at {\em linear order} in the $H$-flux 
the above background  provides a solution.
We can thus conclude that the flat-space background with 
constant $H$ and $\Phi$ corresponds to a 
bona fide conformal field theory at linear order in the flux.
Furthermore,  since the
three-bracket \eqref{result2} is linear in  $\theta^{abc}\sim H^{abc}$,
up to first order in the $H$-flux we expect to find 
a reliable world-sheet CFT framework 
capturing potential non-associative effects.


\paragraph{Three-current correlators}

We continue by noting that the closed string coordinates $X^a(z,\ov z)$ appearing in \eqref{action_730178}
are actually not proper conformal fields. Only the currents  have a well-defined behavior under conformal transformations. Therefore, as usual, for the free theory we define
\eq{
  \label{def_cur_01}
  J^a(z) = i \hspace{0.5pt}\partial X^a(z) \;, \hspace{60pt}
  \ov J{}^a(\ov z) = i \hspace{0.5pt}\ov \partial X^a(z) \;,
}
which at zeroth order in $H$ are indeed holomorphic and anti-holomorphic, respectively.
Employing now the framework of conformal perturbation theory (see \cite{Blumenhagen:2011ph} for more details on the computation), for the  correlators of three currents \eqref{def_cur_01}  (up to first order in the $H$-flux) we find
\eq{
 \bigl\langle J^a(z_1)\, J^b(z_2)\, J^c(z_3) \bigr\rangle  &=  -i\,\frac{{\alpha'}^2}{8}H^{abc}
   \frac{1}{ z_{12}\, z_{23}\, z_{13}} \,, 
   \hspace{18pt}
  \bigl\langle J^a(z_1)\, J^b(z_2)\, \ov J{}^c(\ov z_3) \bigr\rangle =  -i\,\frac{{\alpha'}^2}{8}H^{abc}
   \frac{\ov z_{12}}{ z_{12}^2\, \ov z_{23}\, \ov z_{13}} \,, \\
  \bigl\langle \ov J{}^a(\ov z_1)\, \ov J{}^b(\ov z_2)\, \ov J{}^c(\ov z_3) \bigr\rangle &= +i\,\frac{{\alpha'}^2}{8}H^{abc}   \frac{1}{ \ov z_{12}\, \ov z_{23}\, \ov z_{13}} \,,
   \hspace{18pt}
    \bigl\langle \ov J{}^a(\ov z_1)\, \ov J{}^b(\ov z_2)\, J^c(z_3) \bigr\rangle = +i\,\frac{{\alpha'}^2}{8}
    H^{abc}   \frac{z_{12}}{ \ov z_{12}^2\,  z_{23}\, z_{13}} \,,
}
where we made use of the anti-symmetry of $H_{abc}$, raised the indices of $H$ with $\delta^{ab}$
and used $z_{ij}=z_i-z_j$.
As one can see, these expressions are not purely holomorphic or purely anti-holomorphic, but mixed terms appear. However, we have been using  the currents \eqref{def_cur_01} which are only valid for the free theory. To work at first order in the flux, we should take into account  corrections to \eqref{def_cur_01} linear in $H$. Let us therefore 
define new fields $\mathcal J^a$ and $\ov{\mathcal J}{}^a$ in terms of \eqref{def_cur_01} in the following way
\eq{
  \label{def_cur_04}
  {\cal J}^a (z,\ov z)  = J^a(z)-
  {\textstyle \frac{1}{2}}\hspace{0.5pt}H^a{}_{bc} \, J^b(z) \, 
  {X}^c_R(\ov z)  \;, \hspace{40pt}
  \ov {\cal J}^a (z,\ov z)  = \ov J{}^a(\ov z)-
  {\textstyle \frac{1}{2}}\hspace{0.5pt}H^a{}_{bc} \, X_L^b( z) \,
  \ov J{}^c(\ov z)  \;.  
}
For these, the only non-vanishing correlators of three fields (up to first order in the flux) are then either purely holomorphic or purely anti-holomorphic
\eq{
  \label{cur_cor_19}
 \bigl\langle {\cal J}^a(z_1,\ov z_1)\, {\cal J}^b(z_2,\ov z_2)\, {\cal J}^c(z_3,\ov z_3) \bigr\rangle  
 &=  -i\:\frac{{\alpha'}^2}{8}\,H^{abc}\:
   \frac{1}{ z_{12}\, z_{23}\, z_{13}} \;, \\
  \bigl\langle \ov{\cal J}{}^a(z_1,\ov z_1)\, \ov{\cal J}{}^b(z_2,\ov z_2)\,
  \ov {\cal J}{}^c(z_3,\ov z_3) \bigr\rangle 
  &= +i\:\frac{{\alpha'}^2}{8}\,H^{abc}\:
   \frac{1}{ \ov z_{12}\, \ov z_{23}\, \ov z_{13}} \;.
}
Furthermore, using the equation of motion derived from the action \eqref{action_730178}, at linear order in $H$ we compute
$ \ov\partial \mathcal J^a(z,\ov z) = 0$ and $\partial \ov{\mathcal J}{}^a(z,\ov z) = 0$,
so these fields are indeed holomorphic and anti-holomorphic, respectively. From now on, they
will be denoted as ${\cal J}^a(z)$ and $\ov{\cal J}{}^a(\ov z)$.


\paragraph{Current algebra and energy-momentum tensor}

Let us now study  the fields ${\cal J}^a(z)$ and $\ov{\mathcal J}{}^a(\ov z)$ in more detail. Their non-vanishing two-point function up to first order in  $H$ is readily found to be 
\eq{
  \label{ope_98}
  \arraycolsep2pt
  \begin{array}{ccccc}
  \displaystyle \bigl\langle {\cal J}^a (z_1) {\cal J}^b (z_2) \bigr\rangle 
  &=& 
  \displaystyle \bigl\langle J^a(z_1)\, J^b(z_2) \bigr\rangle_0 
  &=&
  \displaystyle  \frac{\alpha'}{2} \:\frac{1}{(z_1-z_2)^2}\:\delta^{ab} \;, \\[3mm]
  \displaystyle \bigl\langle \ov{\cal J}{}^a (\ov z_1)\, \ov{\cal J}{}^b (\ov z_2) \bigr\rangle 
  &=& 
  \displaystyle \bigl\langle \ov J{}^a(z_1)\, \ov J{}^b(z_2) \bigr\rangle_0 
  &=& 
  \displaystyle \frac{\alpha'}{2} \:\frac{1}{(\ov z_1-\ov z_2)^2}\:\delta^{ab} \;, 
  \end{array}
}
where we employed the definition \eqref{def_cur_01} of the currents $J^a(z)$
as well as the two-point function of the fields $X^a(z,\ov z)$.
Taking then into account the three-point functions  \eqref{cur_cor_19} of the fields $\mathcal J^a(z)$ and $\ov{\mathcal J}{}^a(\ov z)$, with the help of \eqref{ope_98} we can construct the following OPEs 
\eq{
  \label{OPE_01}
  \arraycolsep2pt
  \begin{array}{cclclcl}
  \displaystyle {\cal J}^a(z_1)\; {\cal J}^b(z_2) 
  &=&
  \displaystyle  \frac{\alpha'}{2} \,\frac{\delta^{ab}}{(z_1-z_2)^2}
  &-&
  \displaystyle \frac{\alpha'}{4}\, \frac{i\, H^{ab}{}_c}{z_1-z_2}\: {\cal J}^c(z_2) 
  &+&
  \displaystyle  {\rm reg.} \;, \\[4mm] 
  \displaystyle \ov{\cal J}{}^a(\ov z_1)\; \ov{\cal J}{}^b(\ov z_2) 
  &=& 
  \displaystyle \frac{\alpha'}{2} \,\frac{\delta^{ab}} {(\ov z_1-\ov z_2)^2}
  &+&
  \displaystyle \frac{\alpha'}{4}\, \frac{i\, H^{ab}{}_c}{\ov z_1-\ov z_2}\: \ov{\cal
    J}{}^c(\ov z_2) 
  &+&
  \displaystyle {\rm reg.}\;,
  \end{array}
}
where ``reg.'' stands for regular terms and where the OPEs 
between ${\cal J}^a(z)$ and $\ov{\cal J}{}^b(\ov z)$ are purely regular.
Note that \eqref{OPE_01} defines two independent non-abelian current algebras 
with structure constants $f^{ab}{}_c\simeq H^{ab}{}_c$. The only difference to the usual 
expressions is an opposite relative sign for $H^{ab}{}_c$ between the
holomorphic and anti-holomorphic parts.

Next, we turn to the  energy-momentum tensor. Up to linear order in the flux we find 
\eq{
  \label{def_emt_01}
  {\cal T}(z) = \frac{1}{\alpha'}\: \delta_{ab}:\! {\cal J}^a {\cal J}^b\!:\! (z) \;, \hspace{40pt}
  \ov{\cal T}(\ov z) = \frac{1}{\alpha'}\: \delta_{ab} :\! \ov{\cal J}{}^a\ov{\cal J}{}^b\!:\! (\ov z) \;,
}
and the anti-symmetry of $H$ implies that the OPEs of two energy-mo\-men\-tum tensors take the form
\eq{
  \label{OPE_02}
  \arraycolsep2pt
  \begin{array}{ccccc}
  \displaystyle {\cal T}(z_1)\; {\cal T}(z_2) 
  &=&
  \displaystyle  \frac{{c/2}}{(z_1-z_2)^4} + \frac{2\, {\cal T}(z_2)}{(z_1-z_2)^2}
  + \frac{\partial {\cal T}(z_2)}{z_1-z_2} 
  &+& {\rm reg.} \;, \\[4mm]
  \displaystyle \ov{\cal T}(\ov z_1)\; \ov {\cal T}(\ov z_2) 
  &=&
  \displaystyle  \frac{c/2}{(\ov z_1-\ov
   z_2)^4} + \frac{2\, \ov{\cal T}(z_2)}{(\ov z_1-\ov z_2)^2}
  + \frac{\partial \ov{\cal T}(z_2)}{\ov z_1-\ov z_2} 
  &+& {\rm reg.} \;, 
  \end{array}
}
with $ {\cal T}(z_1) \, \ov {\cal T}(\ov z_2)$ regular.
We therefore find two copies of the  Virasoro algebra with the same central
charge $c$ as for the free theory.
Moreover, using \eqref{OPE_01} and again the anti-symmetry of $H$, one
can show that the fields ${\cal J}^a(z)$ and $\ov{\cal J}{}^a(\ov
z)$ are  primary  of  conformal dimension $(1,0)$ and $(0,1)$ with respect to $\mathcal T(z)$ and $\ov{\mathcal T}(\ov z)$; and so they are indeed non-abelian  currents.


\paragraph{Basic three-point function}

Let us now  define fields ${\cal X}^a(z,\ov z)$ as the integrals of \eqref{def_cur_04}. In particular, we write
\eq{
  \mathcal J^a(z) = i\hspace{0.5pt}\partial \mathcal X^a(z,\ov z) \;,\hspace{60pt}
  \ov{\mathcal J}{}^a(\ov z) = i\hspace{0.5pt}\ov \partial \mathcal X^a(z,\ov z) \;.
}  
The three-point function of three $\mathcal X^a$ up to first order in the $H$-flux can then be obtained by integrating the corresponding  correlators \eqref{cur_cor_19}.
For that purpose, we introduce
the Rogers dilogarithm $L(z)$ which is defined in terms of the usual dilogarithm ${\rm Li}_2(z)$ as follows
\eq{
  L(z)={\rm Li}_2(z) + \frac{1}{ 2} \log (z) \log(1-z)\; .
}
For the correlator of three fields $\mathcal X^a(z,\ov z)$ one  obtains (see \cite{Blumenhagen:2011ph} for more details)
\eq{
  \label{tpf_01}
  &\bigl\langle {\cal X}^a(z_1,\ov z_1)\, {\cal X}^b(z_2,\ov z_2)\,   {\cal X}^{c}(z_3,\ov z_3) \bigr\rangle   
   = \frac{{\alpha'}^2}{12}\: H^{abc} 
   \biggl[ L \Bigl( {\textstyle \frac{z_{12}}{z_{13}} } \Bigr) 
  + L \Bigl( {\textstyle \frac{z_{23}}{z_{21}} } \Bigr) 
  + L \Bigl( {\textstyle \frac{z_{13}}{z_{23}} } \Bigr) - {\rm c.c.} \biggr]
   \;,
}
where ``c.c.'' stands for complex conjugation.
To simplify our notation for the following, let us recall from \eqref{setup_01} the relation between the flux $H$ and the flux parameter $\theta$, that is
$\theta^{abc}=\frac{{\alpha'}^2}{12} H^{abc}$, 
and let us introduce
\begin{equation}
   {\cal L}(z)=L(z)+L\left({\textstyle 1-\frac{1}{z}}\right)
    + L\left({\textstyle \frac{1}{1-z}}\right)\; .
\end{equation}
The correlation function \eqref{tpf_01} of three fields $\mathcal X^a(z,\ov z)$ in the $H$-flux background can then be written as
\begin{equation}
  \label{three-point_01}  
  \bigl\langle {\cal X}^a(z_1,\ov z_1)\, {\cal X}^b(z_2,\ov z_2)\,    {\cal X}^{c}(z_3,\ov z_3) \bigr\rangle 
  = {\theta^{abc}} \Bigl[{\cal L}
   \bigl( {\textstyle \frac{z_{12}}{ z_{13}} }\bigr) - {\cal L}
   \bigl({\textstyle \frac{\ov z_{12}}{ \ov z_{13}}}\bigr) \Bigr]\; .
\end{equation}


\paragraph*{Vertex operator for the tachyon}

We now define vertex operators. In analogy to the free theory of  a closed string without $H$-flux, which in a compact space can  have momentum $p_a$ and winding $w^b$, we define  left- and right-moving momenta $k_{L/R}$ as 
\eq{
  \label{def_mom_76}
  k_{L}^a = p^a + \frac{w^a}{\alpha'} \;, \hspace{60pt}
  k_{R}^a = p^a - \frac{w^a}{\alpha'} \;.
}
The vertex operator for the perturbed theory should then be written in the following way
\eq{
  \label{def_vo7023}
  {\cal V}(z,\ov z) = \, :\!\exp \bigl( i\hspace{0.5pt} k_L\cdot {\cal X}_L +
  i\hspace{0.5pt} k_R \cdot {\cal X}_R \bigr) \!: \;,
}
where we  employ the short-hand notation $k_L\cdot {\cal X}_L= k_{La} {\cal X}^a_L$, $:\ldots :$ denotes normal ordering and 
 the left- and right-moving fields ${\cal X}^a_{L/R}$ are obtained via
integration of  the currents.
Now, recall that in the free theory the tachyon vertex operator
is a primary field  of conformal dimension $(h,\ov h)=(\frac{\alpha'}{4}\,
k_L^2,\frac{\alpha'}{4}\, k_R^2)$, and
in covariant quantization of the bosonic string
physical states are given by primary fields of conformal
dimension $(h,\ov h)=(1,1)$. 
In the deformed theory, we also require that vertex operators ${\cal V}(z,\ov z)$ 
are primary with respect to ${\cal T}(z)$ and 
$\ov {\cal T}(\ov z)$ which
is not guaranteed a priori.
However, it is again the anti-symmetry of $H$ which implies 
\eq{
   \arraycolsep2pt
   \begin{array}{lclclcl}
   \displaystyle {\cal T}(z_1)\, {\cal V}(z_2,\ov z_2) 
   &=& 
   \displaystyle \frac{1}{(z_1-z_2)^2} \, \frac{\alpha' k_L\cdot k_L}{4} \,   {\cal V}(z_2,\ov z_2)
   &+&
   \displaystyle \frac{1}{z_1-z_2} \, \partial {\cal V}(z_2,\ov z_2) 
   &+&{\rm reg.} \;, \\[3mm]
   \displaystyle \ov{\cal T}(\ov z_1)\, {\cal V}(z_2,\ov z_2) 
   &=& 
   \displaystyle \frac{1}{(\ov z_1-\ov z_2)^2} \, \frac{\alpha' k_R\cdot k_R}{4} \,  {\cal V}(z_2,\ov z_2)
   &+&
   \displaystyle \frac{1}{\ov z_1-\ov z_2} \, \ov\partial {\cal V}(z_2,\ov z_2) 
   &+&{\rm reg.} \;.
   \end{array}  
}
Thus,  the vertex operator \eqref{def_vo7023} is  primary and 
has the correct conformal dimension, and is therefore a physical quantum
state of the  deformed theory.
However, as it is explained in more detail in 
\cite{Blumenhagen:2011ph},
in order for the tachyon vertex operator to carry momenta $(k_L,k_R)$ we have to require
\eq{
  \label{M1860}
  0= H^a{}_{bc}\, k_L^b\, k_R^c \simeq
  H^a{}_{bc}\, p^b\, w^c  \simeq  \bigl[\, \vec p \times \vec w \,\bigr]^a\;.
}


\subsection{T-duality}
\label{sec_Tdual}

As expected from the string equations of motion \eqref{targeteom}, in the last subsection we 
have found a conformal field theory which
describes the sigma model for a flat metric and constant $H$-flux up to
linear order. However, we are also interested in backgrounds T-dual to the $H$-flux configuration.
On the level of the CFT, T-duality is usually realized as
a reflection on the right-moving coordinates. Since 
the corrected fields ${\cal X}^a(z,\ov z)$  still admit
a split into a holomorphic and an anti-holomorphic piece,
we define T-duality on the  world-sheet  along a direction $\mathcal X^a$ as 
\eq{
  \begin{array}{c}
  {\cal X}_L^a(z) \\[1mm]
  {\cal  X}_R^a(\ov z)   
  \end{array}
  \qquad
  \xrightarrow{\;\mbox{\scriptsize T-duality}\;}
  \qquad
  \begin{array}{c}
  +{\cal X}_L^a(z)\;, \\[1mm]
  -{\cal  X}_R^a(\ov z) \;.
  \end{array}
}
Clearly, for the currents this implies a similar action, 
and so  the ``structure constants'' $H^{ab}{}_c$ 
in the  anti-holomorphic OPE \eqref{OPE_01} receive an additional minus sign
when performing a T-duality transformation.

In the next subsection, we compute scattering amplitudes
for tachyon vertex operators in the $H$-flux background.
There we allow for  momentum and winding along the 
directions of our three-dimensional (compact) space specified by \eqref{setup_01}.
However, in the T-dual  backgrounds  we are particularly interested in pure momentum
scattering, as from there one would derive the
low-energy effective action as a derivative expansion.
Now, above we have mentioned that a tachyon vertex operator $\mathcal V(z,\ov z)$
indeed corresponds to a physical state provided that 
 $ \vec p \times \vec w =\vec 0$.
This has the following implications (see again \cite{Blumenhagen:2011ph} for further details):

\begin{itemize}

\item The effective field theory 
for tachyons in the $H$-flux
background is expected to be reliably computable 
from scattering amplitudes of pure momentum tachyons, since in this case
the constraint \eqref{M1860} is satisfied.

\item For backgrounds with geometric flux, pure momentum scattering is related 
to the scattering of for instance $(p_1,p_2,\omega_3)$ modes in the $H$-flux background. However, in this case $\vec \omega\times\vec p \neq \vec 0$ and so we
can not reliably employ the vertex operator for the tachyon in our present approach.
A similar situation occurs for pure momentum scattering in the $Q$-flux background.

\item For the case
of $R$-flux we can again reliably compute the scattering
amplitudes for pure momentum tachyons. By T-duality, they  are   related
to the scattering of pure winding states in
the $H$-flux background
for which \eqref{M1860} is satisfied.

\end{itemize}
Therefore, in the following we focus on backgrounds with $H$- and $R$-flux.


\subsection{Tachyon scattering amplitudes}
\label{sec_threep}

In this subsection, we compute  scattering amplitudes 
of the tachyon vertex operators \eqref{def_vo7023} and extract a tri-product.


\paragraph{Three-tachyon amplitude}

As discussed above, we focus on a compact three-dimensional space and therein
we are interested in pure momentum $(p_1,p_2,p_3)$ or pure winding $(w_1,w_2,w_3)$ state scattering, where the latter is related by three T-dualities to pure momentum scattering in the $R$-flux
background. We therefore consider vertex operators of form
\eq{
  \label{vo_tach_19}
  \arraycolsep2pt
  \begin{array}{lclcl}
  {\cal V}_i^H &\equiv& 
  {\cal V}_{p_i}(z_i,\ov z_i) &=& 
  \displaystyle :\! \exp \bigr(\, i\hspace{0.5pt} p_i \cdot {\cal X}(z_i,\ov z_i) \bigl) : \;, \\[1.75mm]
  {\cal V}_i^R &\equiv& 
  {\cal V}_{w_i}(z_i,\ov z_i) &=& 
  \displaystyle :\! \exp \bigr(\, i\hspace{0.5pt} w_i \cdot \widetilde{\cal X}(z_i,\ov z_i) \bigl) : \;,
  \end{array}
}
where $\widetilde{\cal X}={\cal X}_L-{\cal X}_R$.  Note that here and in the following
we  employ the short hand notation ${\cal V}^{H/R}_i$ and $V^{H/R}_i$
for the vertex operators of the perturbed and free theory, respectively.
Furthermore, since we can consider  ${\cal V}_i^R$ as the momentum vertex operator
in the  $R$-flux background, in the following we set $w|_H\to p|_R$.

Next, in string theory one has to work with vertex operators integrated over the world-sheet. We therefore define 
\eq{ 
   \label{some_def_845}
   \mathcal T^{H/R}_{i} = \int d^2z\,   {\cal V}^{H/R}_{i} \; .
 }  
Taking into account the freedom to fix three points on the world-sheet via the $SL(2,\mathbb C)$ symmetry, the three-tachyon scattering amplitude is  then given by
\eq{
  \label{threetachyon}
  &\bigl\langle\, \mathcal T_1\: \mathcal T_2\:\mathcal  T_3\, \bigr\rangle^{H/R} 
  =  \int  \prod_{i=1}^3 d^2 z_i\: \delta^{(2)}(z_i-z_i^0)\, \vert
   z_{12}\,z_{13}\,z_{23}\vert^2  \, 
    \bigl\langle \,{\cal V}_1\, {\cal V}_2 \,{\cal V}_3\, \bigr\rangle^{H/R} \;,
}
where we have put the superscript indicating the $H$- and $R$-flux background
outside the bracket in order to shorten the notation.
Using \eqref{three-point_01}, for the correlator of three vertex operators one obtains
\eq{
  \label{threetachyonb}
   \bigl\langle \,{\cal V}_1 \,{\cal V}_2 \,{\cal V}_3 \,\bigr\rangle^{H/R}
   =\frac{\delta(p_1+p_2+p_3)}{\vert z_{12}\,z_{13}\,z_{23}\vert^2}
       \exp\Bigl[ -i\hspace{0.5pt}\theta^{abc}\, p_{1,a} p_{2,b} p_{3,c}  \bigl[
  {\cal L}   \bigl( {\textstyle \frac{z_{12}}{z_{13}} }\bigr) \mp {\cal L}
   \bigl({\textstyle \frac{\ov z_{12}}{ \ov z_{13}}}\bigr)\bigr] \Bigr]_{\theta} \,,
}
where $[\ldots]_{\theta}$ indicates that the result is valid only up to linear order in $\theta$. Note that the upper sign corresponds to the case of $H$-flux and the lower to the case of $R$-flux.
The full scattering amplitude then becomes
\eq{
  \label{threetacyon}
  &\bigl\langle\, \mathcal T_1\: \mathcal T_2\:\mathcal  T_3\, \bigr\rangle^{H/R}
  = \int \prod_{i=1}^3  d^2 z_i\, \delta^{(2)}(z_i-z_i^0) \, \delta(p_1+p_2+p_3) \times \\
 & \hspace{140pt}
 \exp\Bigl[ -i\hspace{0.5pt}\theta^{abc}\, p_{1,a} p_{2,b} p_{3,c}  \bigl[
  {\cal L}   \bigl( {\textstyle \frac{z_{12}}{z_{13}} }\bigr) \mp {\cal L}
   \bigl({\textstyle \frac{\ov z_{12}}{ \ov z_{13}}}\bigr)\bigr] \Bigr]_{\theta}.
}


\paragraph{Permutations}

Let us now study the behavior of \eqref{threetachyonb}  under
permutations of the vertex operators ${\cal V}^{H/R}_i$. 
Before applying momentum conservation, 
the three-tachyon amplitude for a permutation $\sigma$  can be computed using properties of the Rogers dilogarithm.
With $\epsilon=-1$ for  $H$-flux and $\epsilon=+1$ for the case of $R$-flux, 
one finds
\begin{equation}
  \label{phasethreeperm}
  \bigl\langle \, {\cal V}_{\sigma(1)}   {\cal V}_{\sigma(2)}  {\cal V}_{\sigma(3)}  \bigr\rangle^\epsilon=
  \exp\Bigl[ \,i\left({\textstyle \frac{1+\epsilon}{ 2}}\right)  \eta_\sigma\,  \pi^2\,  \theta^{abc}\, p_{1,a} 
  \,p_{2,b} \,p_{3,c} \Bigr]
  \bigl\langle {\cal V}_1\,  {\cal V}_2\,  {\cal V}_3  \bigr\rangle^\epsilon \;,
\end{equation}
where in addition $\eta_{\sigma}=1$ for an odd permutation and 
$\eta_{\sigma}=0$ for an even one. 
Thus,  for the $R$-flux background a non-trivial phase may appear
which we have established up to linear order in the flux. 
As will be discussed in more detail in section \ref{secthree}, 
the phase in \eqref{phasethreeperm} can be recovered from 
a three-product on the space of  functions
$V_{p_n}(x)=\exp( i\, p_n \cdot x )$, which can be defined as 
\eq{
  \label{threebracketexp}
   V_{p_1}(x)\,\tri\, V_{p_2}(x)\, \tri\, V_{p_3}(x)\stackrel{\rm def}{=} 
    \exp\Bigl( -i \,{\textstyle \frac{\pi^2}{2}}\, \theta^{abc}\,
   p_{1,a}\, p_{2,b}\, p_{3,c} \Bigr) V_{p_1+p_2+p_3}(x)\; .
}
However, in correlation functions operators
are understood to be radially ordered and so changing the order
of operators should not change the form of the amplitude.
This is known as crossing symmetry which is one of the defining
properties of a CFT and thus should also be satisfied
here. In the case of the $R$-flux background, this
is reconciled by applying momentum conservation leading to
\eq{
  p_{1,a} \,p_{2,b}\, p_{3,c}\,\theta^{abc} = 0
  \hspace{40pt}{\rm for}\hspace{40pt} p_3=-p_1-p_2 \;.
}
Therefore, scattering amplitudes of three tachyons do not receive
any corrections at linear order in $\theta$  both for the $H$- and
 $R$-flux
\eq{
  \bigl\langle\, \mathcal T_1\: \mathcal T_2\:\mathcal  T_3\,
  \bigr\rangle^{H/R}=\delta(p_1+p_2+p_3) \; .
}


\paragraph{Four- and N-tachyon amplitudes}

We now want to detect phases possibly appearing for the product of $N$ closed string
tachyon vertex operators. But before we consider the general case,
let us start with the amplitude of four tachyons.
Up to linear
order in $\theta$ we obtain
\eq{
\label{fourptampl}
  &\bigl\langle {\cal V}_1\,{\cal V}_2\,{\cal V}_3\,{\cal V}_4 \bigr\rangle^{H/R} = 
  \bigl\langle V_1\,V_2\,V_3\,V_4 \bigr\rangle^{H/R}_0 \; \exp \biggl[  -i \hspace{0.5pt}
   \theta^{abc}\!\!\! \sum_{1\leq i<j<k\leq 4}  p_{i ,a} \,p_{j,b}\, p_{k,c}  
   \Bigl[ {\cal L}\bigl({\textstyle \frac{z_{ij}}{ z_{ik}}}\bigr)
  \mp  {\cal L}\bigl({\textstyle\frac{\ov z_{ij}}{ \ov z_{ik}}}\bigr)
  \Bigr] \biggr]_{\theta} .
}
Again, the difference between $H$- and $R$-flux is
given by the sign between the holomorphic and  the anti-holomorphic contribution, and 
the four-point function $\bigl\langle V_1\,V_2\,V_3\,V_4 \bigr\rangle^{H/R}_0$
is just the one from the free theory.
We can now  determine the behavior of the amplitude under a permutation 
of  the vertex operators. Prior to using momentum conservation,  we 
again find momentum dependent phase factors. 
Analogous to the three-tachyon amplitude, these arise in the case of $R$-flux 
and  can be described as resulting from a deformed
four-product of the form
\eq{
  \label{fourbracketexp}
   &V_{p_1}(x)\,\tri_4\, V_{p_2}(x)\, \tri_4\, V_{p_3}(x) \, \tri_4\, V_{p_4}(x)\stackrel{\rm def}{=} 
   \exp\Bigl[-i\, {\textstyle \frac{\pi^2}{2}} \, \theta^{abc}\, \bigl(
   p_{1,a}\, p_{2,b}\, p_{3,c} \\
   &\hspace{100pt}
   +p_{1,a}\, p_{2,b}\, p_{4,c} + 
   p_{1,a}\, p_{3,b}\, p_{4,c} +
   p_{2,a}\, p_{3,b}\, p_{4,c}
   \bigr) \Bigr]\, V_{\sum p_i}(x)\; .
}
However, employing momentum conservation, one can show that this phase
becomes trivial so that the four-tachyon amplitude is indeed
 crossing symmetric.

This computation for four tachyons can straightforwardly be generalized to higher $N$-tachyon
amplitudes.
The phase factors appearing when permuting two vertex operators for the case of the $R$-flux
background can then be encoded in a deformed $N$-product  of the form
\eq{
\label{Nbracketexp}
   V_{p_1}(x)\,\tri_N\, \ldots\, \tri_N\, V_{p_N}(x) \stackrel{\rm def}{=} 
   \exp\Bigl(-i \,{\textstyle \frac{\pi^2}{2}}\, \theta^{abc}\!\!\!
   \sum_{1\le i < j <  k\le N} \!\!
   p_{i,a}\, p_{j,b}\, p_{k,c}   \Bigr)\; V_{\sum p_i}(x)\; .
}
The phase becomes again trivial after employing momentum conversation so that
all $N$-tachyon correlators are crossing symmetric. 
This signals that the basic principle of perturbative closed string theory,
namely conformal field theory, seems to be compatible with  
non-geometric backgrounds for which the $N$-product of functions
is deformed by \eqref{Nbracketexp}.


\subsection{A tri-product}
\label{secthree}

In this section, we show that the relative phase factors
of tachyon correlation functions
can be rephrased in terms of a generalization of the
Moyal-Weyl star-product, which we call a tri-product.


\paragraph{Tri-product}

In particular, the phase appearing in the three-point correlator \eqref{phasethreeperm}
indicates that we can define a three-product of functions
$f(x)$ in the following way
\eq{
\label{threebracketcon}
\bom{
   f_1(x)\,\tri\, f_2(x)\, \tri\, f_3(x) \stackrel{\rm def}{=} \exp\Bigl(
   {\textstyle \frac{\pi^2}{ 2}}\, \theta^{abc}\,
      \partial^{x_1}_{a}\,\partial^{x_2}_{b}\,\partial^{x_3}_{c} \Bigr)\, f_1(x_1)\, f_2(x_2)\,
   f_3(x_3)\Bigr|_{x} \;,
   }
}
where we used the notation $(\ )\vert_x=(\ )\vert_{x_1=x_2=x_3=x}$.
Choosing $f_n(x)=\exp( i\hspace{0.5pt} p_n \cdot x )$ we obtain formula
\eqref{threebracketexp},
which after integration over $x$ gives
\eq{
\label{threebracketexpint}
   \int d^3x\;  f_1(x)\,\tri\, f_2(x)\, \tri\, f_3(x) &=  \exp\Bigl(-i {\textstyle \frac{\pi^2}{ 2}}\theta^{abc}\,
   p_{1,a}\, p_{2,b}\, p_{3,c} \Bigr)\,  \delta(p_1+p_2+p_3)\\
   &=\int d^3x\; f_1(x)\, f_2(x)\,  f_3(x)\; .
}
Note that \eqref{threebracketcon} is precisely the three-product \eqref{result2}. 
Indeed, 
with $P^3$ denoting the permutation group of three elements,
the three-bracket for the coordinates $x^a$ can then be re-derived  as
the completely anti-symmetrized sum of three-products
\eq{
\label{antisymtripcon}
   \bigl [x^a,x^b,x^c \bigr]=\sum_{\sigma\in P^3} {\rm sign}(\sigma) \;  
     x^{\sigma(a)}\, \tri\,  x^{\sigma(b)}\, \tri\,  x^{\sigma(c)} =
     3\pi^2\, \theta^{abc}\; .
}


\paragraph{N-product}

Next, we consider the $N$-tachyon amplitude and the phase appearing
in equation \eqref{Nbracketexp}. This motivates us to define
the $N$-product
\eq{
   f_1(x)\, \tri_N\,  f_2(x)\, \tri_N \ldots \tri_N\,  f_N(x) \stackrel{\rm def}{=} 
   \exp\left[ {\textstyle \frac{\pi^2}{2}} \theta^{abc}\!\!\!\!\! \sum_{1\le i< j < k\le N}
     \!\!\!\!  \, 
      \partial^{x_i}_{a}\,\partial^{x_j}_{b} \partial^{x_k}_{c} \right]\, 
   f_1(x_1)\, f_2(x_2)\ldots
   f_N(x_N)\Bigr|_{x} ,
}
which is the closed string generalization of the open string
non-commutative product \eqref{Nbracketcon}.
This completely defines the new tri-product, which satisfies
the relation
\eq{
\label{thenicerel}
    f_1\,\tri_N\,  f_2  \,\tri_N\,  \ldots\,  \tri_N\, f_{N-1} \, \tri_N\, 1\,   \
   =  f_1\,\tri_{N-1}\,  \ldots\,  \tri_{N-1}\,  f_{N-1} \;.
} 
Specializing this expression  to $N=3$ gives
\begin{equation}
    f_1\,  \tri_2\,  f_2 = f_1\,\tri_3\, f_2\,\tri_3 \, 1= f_1 \cdot f_2  \; , 
\end{equation}
which just means that the tri-product of two functions
is the usual commutative point-wise product.
However, there are two main differences compared to the open-string case.
\begin{itemize}

\item For the open string the star $N$-product was 
related to successive application of the usual Moyal-Weyl
bi-product.  
This simplifying behavior is not true for the tri-product, i.e.
the $N$-products $\ \tri_N$ {\it can not} be related to successive
applications of the three-product $\ \tri\;\,=\;\,\tri_3$. 
This means that one does not only have to specify a deformed product of three functions
(with the rest following), but has to specify a
definition for a deformed product of any number of functions.

\item In contrast to the open string case, the effect
of the tri-product in integrals vanishes, i.e. 
\eq{
\int d^nx\, f_1(x)\, \tri_N\,  f_2(x)\, \tri_N \ldots \tri_N\,  f_N(x)
=  \int d^nx\, f_1(x)\,  f_2(x)\,  \ldots   f_N(x)\; .
}
In other words, the difference between the tri-product and
the ordinary product  is a total derivative.
Thus, closed strings can consistently
be defined on non-associative backgrounds since  in string scattering amplitudes
its effect vanishes.

\end{itemize}


\clearpage
\section{Summary}

In these proceedings, we have summarized recent work on non-commutativity in closed string theory which appeared in \cite{Blumenhagen:2010hj}  and  \cite{Blumenhagen:2011ph}. In particular, motivated by results from the open-string sector, we have illustrated two approaches to study non-commutative behavior  for closed strings.

The first approach mimics the computation of the commutator of two open-string coordinates. However, in the case of a closed string one should  consider a cyclic double commutator involving three  instead of two fields. If such an expression is non-vanishing, it not only indicates a non-commutative but also a non-associative structure. Furthermore, we illustrated that in order to obtain a non-trivial result, background fluxes have to be non-vanishing. These can be ordinary $H$-fluxes or geometric fluxes, but also non-geometric $Q$- and $R$-fluxes. In section \ref{sec_3bracket}, we have computed the cyclic double commutator shown in equation \eqref{result2} and found it to be non-vanishing for a background with $R$-flux, and therefore indicating a non-associative structure.

The second approach to non-commutativity considered here is to compute correlation functions of vertex operators. Since the non-commutative effect obtained in \eqref{result2} is linear  in the flux parameter, we studied a conformal field theory up to linear order in the background flux. Defining then a tachyon vertex operator in this CFT and computing correlation functions thereof, we obtained a phase factor which encodes the flux dependence (up to linear order). From this factor we motivated and studied a tri-product \eqref{threebracketcon} which captures  non-commutative and non-associative effects.

The direction for future work is to make the non-commutative and non-associative structure more apparent. For instance, the relation between $R$-flux and the non-vanishing three-bracket \eqref{result2} leading to a non-associative behavior has to be understood better. Furthermore, properties of the tri-product \eqref{threebracketcon} have to be studied, which is needed to eventually formulate an effective theory using that product.

\vskip2cm

\paragraph{Acknowledgements}
We would like thank Ralph Blumenhagen, Andreas Deser, Dieter L\"ust and Felix Rennecke for collaboration and  discussions on the material presented in this talk. 
We are also grateful to the organizers of the 
``Workshop on Noncommutative Field Theory and Gravity'' at the 
Corfu Summer Institute for the opportunity to present this material.
The author is supported by the Netherlands Organization for Scientific Research (NWO) under the VICI grant 680-47-603.


\clearpage
\nocite{*}
\bibliography{references}  
\bibliographystyle{utphys}


\end{document}